# Theoretical investigation of pseudo-gap state of YBCO

Partha Goswami

*Department of Physics, D.B. College, University of Delhi, Kalkaji,Delhi-110019,India*

**ABSTRACT.** A momentum space, mean field d-density wave (DDW) Hamiltonian is investigated self-consistently. The pseudo-gapped(PG)state of YBCO is assumed to correspond to the pure DDW state. A relation between thermodynamic potential of the system and certain spectral weight functions is established. This yields an expression for entropy in DDW state in the absence of magnetic field. The relation is useful for deriving finite temperature thermodynamics of the system. We show that the PG transition is a first order one and the entropy per unit cell increases in this state. We also analyze the fermion occupancy, in the presence of magnetic field, at the anti-nodal points of the Fermi surface to estimate the frequency of quantum oscillations.

## MAIN TEXT

The high temperature superconductivity in hole-doped cuprates is derived from doping the parent anti-ferromagnetic, charge-transfer insulators. It is now a common view [1,2,3,4] that the partial gap in the normal state and the superconducting (SC) gap in such systems have robust d-wave symmetry. Therefore, in these systems sharp, long-lived quasi-particle like excitations (QPE) remain possible near the nodal region centered around $(\pm\pi/2, \pm\pi/2)$ where the gaps are zero. In the anti-nodal sector centered around $[(\pm\pi, 0), (0,\pm\pi)]$, on the other hand, QPEs are inconspicuous. This is signaled by the broadening of the QPE peak in the spectral function and decrease in their life-time even in the normal state . In fact, the normal state properties of cuprates are highly anomalous, particularly, in the under-doped region. The undoped cuprates are two-dimensional Mott insulators with a large anti-ferromagnetic exchange interaction. In $YBa_2Cu_3O_{7-\delta}$, for example, upon hole doping the anti-ferromagnetism is destroyed at the hole density $\delta \sim 0.10$ and the superconductivity is optimized at $\delta \sim 0.16$. Further doping leads to decreasing $T_c$ and more or less conventional Fermi-liquid behavior. The under-doped metallic region between $0.10 \leq \delta \leq 0.16$ for this system has attracted much attention because anomalies in many physical properties, yielded by the specific heat, magnetic susceptibility, transport, and optical measurements, have led to a widespread belief that therein lies the key to understanding high temperature superconductivity. These measurements suggest a partial gapping of the Fermi surface that has been termed a pseudo-gap[3,4]. Chakravarty et al.[5,6,7] had put forward interpretation of this gap in terms of a hidden long-range order, viz. d-density wave(DDW) order. It breaks symmetries signifying time reversal, translation by a lattice spacing, and a rotation by an angle $\pi/2$, while the product of any two symmetry operations is preserved. As noted by these authors, that the commensurate DDW order doubles the unit cell of the real space lattice because the translational symmetry corresponding to a displacement by the lattice spacing *a* of the square planar CuO lattice is broken. As a result, the conventional Brillouin zone (BZ) in the reciprocal space is halved, or reduced, which is known as the reduced Brillouin zone (RBZ). This unit cell doubling, but without a conventional spin or charge density wave order, plays a crucial role in an analysis involving DDW order. According to these authors PG is a consequence of the competition between DDW and d-wave superconductivity(DSC). Furthermore, there is a possibility of the coexistence of these two orders in the under-doped state of cuprates.

In the present communication our main aim to derive an expression for entropy in DDW state in the absence of magnetic field. The relation is useful for obtaining finite temperature thermodynamics of the system. We show that the pseudo-gap(PG) transition is a first order one and the entropy per unit cell increases in the DDW state. We also estimate the frequency of quantum oscillations analyzing the fermion occupancy in the presence of magnetic field at the anti-nodal points located at electron pockets of the Fermi surface of the system.

We start with the DDW mean field Hamiltonian ($H_{DDW}$) in momentum space in the absence of magnetic field for the normal (pseudo-gapped) state. In the second-quantized notation, including terms corresponding to bi-layer splitting, the Hamiltonian of the system(with index i = (1,2) below corresponding to the two layers) $H_{DDW}$ can be expressed as

$$H_{DDW} = \sum_{k\sigma, i=1,2} [\varepsilon_k \, d^{(i)\dagger}_{k,\sigma} d^{(i)}_{k,\sigma} + \varepsilon_{k+Q} \, d^{(i)\dagger}_{k+Q,\sigma} d^{(i)}_{k+Q,\sigma}]$$

$$+ \sum_{k\sigma, i=1,2} [(i\Delta_k) \, d^{(i)\dagger}_{k,\sigma} d^{(i)}_{k+Q,\sigma} - (i\Delta_k) \, d^{(i)\dagger}_{k+Q,\sigma} d^{(i)}_{k,\sigma}]$$

$$+ \sum_{k\sigma} [t_\perp (k) \, d^{(1)\dagger}_{k,\sigma} d^{(2)}_{k,\sigma} + t_\perp (k) \, d^{(2)\dagger}_{k,\sigma} d^{(1)}_{k,\sigma} + t_\perp (k+Q) \, d^{(1)\dagger}_{k+Q,\sigma} d^{(2)}_{k+Q,\sigma}$$

$$+ t_\perp (k+Q) \, d^{(2)\dagger}_{k+Q,\sigma} d^{(1)}_{k+Q,\sigma}]. \quad (1)$$

Here $\varepsilon_k = \varepsilon_k^{(1)} + \varepsilon_k^{(2)} + \varepsilon_k^{(3)} - \mu$, $\varepsilon_k^{(1)} = -2t \, (c_x + c_y)$, $\varepsilon_k^{(2)} = 4t' \, c_x c_y$, $\varepsilon_k^{(3)} = -2t'' \, (c'_x + c'_y)$, $c_i = \cos k_i a$, $c'_i = \cos 2k_i a$ ( i = x,y), $\Delta_k = (\Delta_0 (T)/2) (\cos k_x a - \cos k_y a)$, and 'a' is the lattice constant. The energy gap $\Delta_k$ is the pseudo gap with $d_{x^2-y^2}$ symmetry. Its origin lies in the pair hopping process. The quantity $\varepsilon_k$ is the normal state tight-binding energy dispersion with t, t', t'' being the hopping elements between nearest, next-nearest(NN) and NNN neighbors, respectively, and µ is the chemical potential. The energy $\varepsilon^{(1)}(\mathbf{k})$ satisfies the perfect nesting condition $\varepsilon^{(1)}(\mathbf{k+Q}) = -\varepsilon^{(1)}(\mathbf{k})$ with ordering wave vector $\mathbf{Q} = (\pm\pi,\pm\pi)$. The effect of bi-layer splitting in YBa$_2$Cu$_3$O$_{7-\delta}$ (YBCO) is given by the parametrization in terms of a momentum conserving tunneling matrix element which for the tetragonal structure corresponds to $t_\perp(k) = (t_\perp/4)(\cos k_x a - \cos k_y a)^2$. In the normal state, from Eq.(1), it is known that the energy eigenvalues (with $j = \pm 1$ ($j = +1$ corresponds to the upper branch(U) and $j = -1$ to the lower branch(L), and $\nu = \pm 1$) are $E_k^{(j,\nu)}(k) = \varepsilon_k^U + \nu\, t_\perp(k) + j[(\varepsilon_k^L)^2 + \Delta_k^2]^{1/2}$ where $\varepsilon_k^U = (\varepsilon_k + \varepsilon_{k+Q})/2$ and $\varepsilon_k^L = (\varepsilon_k - \varepsilon_{k+Q})/2$. We also find in the Fermi surface reconstruction exercise that the matrix element $t_\perp(k)$ corresponding to the bi-layer splitting is not of much significance for the hole pocket; the electron pockets, however, appear to split slightly[see ref.7 and Fig.1]. According to Luttinger sum rule (LSR), for δ away from half-filling ( δ > 0 for hole doping and δ < 0 for electron doping), we have

$$(1+\delta) = (N_s)^{-1} \sum_{k\sigma} [u_k^{+2}(\exp(\beta E_k^{U+})+1)^{-1} + u_k^{-2}(\exp(\beta E_k^{U-})+1)^{-1}$$

$$+ v_k^{+2}(\exp(\beta E_k^{L+})+1)^{-1} + v_k^{-2}(\exp(\beta E_k^{L-})+1)^{-1}] \quad (2)$$

where $N_s$ is the number of unit cells and $\beta = (k_BT)^{-1}$. The Bogolubov coherence factors are given by

$$u_k^{+2} = (1/4)[\,1+ (\varepsilon_k^L/((\varepsilon_k^L)^2 + \Delta_k^2)^{1/2} + t_\perp(k)))\{1 + (t_\perp(k)/((\varepsilon_k^L)^2 + \Delta_k^2)^{1/2})\}],$$

$$u_k^{-2} = (1/4)[\,1+ (\varepsilon_k^L/((\varepsilon_k^L)^2 + \Delta_k^2)^{1/2} + t_\perp(k)))\{1 - (t_\perp(k)/((\varepsilon_k^L)^2 + \Delta_k^2)^{1/2})\}],$$

$$v_k^{+2} = (1/4)[\,1- (\varepsilon_k^L/((\varepsilon_k^L)^2 + \Delta_k^2)^{1/2} + t_\perp(k)))\{1 - (t_\perp(k)/((\varepsilon_k^L)^2 + \Delta_k^2)^{1/2})\}],$$

$$v_k^{-2} = (1/4)[\,1- (\varepsilon_k^L/((\varepsilon_k^L)^2 + \Delta_k^2)^{1/2} + t_\perp(k)))\{1 + (t_\perp(k)/((\varepsilon_k^L)^2 + \Delta_k^2)^{1/2})\}].(3)$$

The gap parameter $\Delta_0(T)$ is given by the equation

$$\Delta_0(T) = \sum_k \{(g/2)(\cos k_x a - \cos k_y a)/[(\varepsilon_k^L)^2 + \Delta_k^2]^{1/2}\}$$

$$\times ((\exp(\beta E_k^{L-})+1)^{-1} - (\exp(\beta E_k^{U+})+1)^{-1})$$

$$+\sum_k \{(g/2)(\cos k_x a - \cos k_y a)/[(\varepsilon_k^L)^2 + \Delta_k^2]^{1/2}\}[1-\{2t_\perp(k)/((\varepsilon_k^L)^2 + \Delta_k^2)^{1/2} + t_\perp(k))\}]$$

$$\times ((\exp(\beta E_k^{L+})+1)^{-1} - (\exp(\beta E_k^{U-})+1)^{-1}). \qquad (4)$$

Equations (2) and (4) are consistent with (1). The quantity g is the pair-hopping amplitude. The parameters we choose (see ref.[6,7]) for the analysis at 10% doping are: t = 0.25 eV, t´ = 0.4t, t´´ = 0.0444 t, and $t_\perp$ = 0.032 t. With these choices of the parameters, and the hole doping of 10%, the chemical potential μ is found to be −0.27 eV. The limits of momentum integrals above are obtained from the FS reconstruction exercise alluded to above (see Fig.1). At this doping level, ignoring the bi-layer splitting, it is quite simple to work out the value of the pseudo-gap temperature T* by performing momentum integrations numerically. We find T* ~ 150 K. As regards the parameter g, it can be adjusted to yield the experimental value $\Delta_0(T) = 0.0825$ eV for T < T*.

The single-particle spectral function (SF) in the spin-σ channel (i.e. the one corresponding to $G^{\sigma,\sigma}(k,\tau) = -\langle T\{d_{k,\sigma}(\tau) d^\dagger_{k,\sigma}(0)\}\rangle$) is given by $A^{\sigma,\sigma}(k,\omega) = (-\pi^{-1})\text{Im } G^{(R)\sigma,\sigma}(k,\omega)$, where $G^{(R)\sigma,\sigma}(k,\omega)$ is a retarded Green's function given by $G^{(R)\sigma,\sigma}(k,\omega) = \int_{-\infty}^{\infty}(d\omega'/2\pi)\{\zeta^{(R)\sigma,\sigma}(k,\omega')/(\omega - \omega' + i 0^+)\}$ and $\zeta^{(R)\sigma,\sigma}(k,\omega) = (1/2\pi i)\{G^{\sigma,\sigma}(k,z)|_{z=\omega-i0+} - G^{\sigma,\sigma}(k,z)|_{z=\omega+i0+}\}$. The spectral functions provide information about the nature of the allowed electronic states, regardless whether they are occupied or not. In the analysis of the spectral function ( $A([(\pm\pi, 0), (0,\pm\pi)], \omega)$ ) at the anti-nodal point, obtained from above at the 10% doping level, we do notice the bi-layer splitting of the main frequencies:

$$A([(\pm\pi, 0), (0,\pm\pi)], \omega) = (1/4\pi)([(\acute{\Gamma}/t)^2/\{(\omega/t)+0.3456)^2 + (\acute{\Gamma}/t)^2\}]$$

$$+ [(\acute{\Gamma}/t)^2/\{(\omega/t)+0.4096)^2 + (\acute{\Gamma}/t)^2\}] + [(\acute{\Gamma}/t)^2/\{(\omega/t)+0.9856)^2 + (\acute{\Gamma}/t)^2\}]$$

$$+[ (\acute{\Gamma}/t)^2 / \{(\omega/t)+1.0496)^2 + (\acute{\Gamma}/t)^2\}]). \qquad (5)$$

In Eq.(5) the delta functions are replaced by Lorenztians. The quasi-particle lifetime is infinite in HFA as the Hamiltonian is quadratic in fields and therefore exactly diagonalizable. As a consequence the corresponding eigen states are stationary states with infinite lifetime. Therefore, in generating the plot (not shown) of Eq.(5), a small artificial broadening of the single particle energies $(\acute{\Gamma}/t) = 0.03\text{-}0.05$ may be assumed. We next consider a retarded Green's function $G^R_\sigma (k,\omega) = (-i) \int_{-\infty}^{\infty} dt \exp(i\omega t)\langle\{d_{k,\sigma}(t), d^\dagger_{k,\sigma}(0)\}\rangle \theta(t)$. The density of states (DOS) in the spin-$\sigma$ channel ($\rho^{\sigma,\sigma}(k,\omega)$) is given by $\rho^{\sigma,\sigma}(k,\omega) = (-\pi^{-1})\text{Im } G^R_\sigma (k,\omega)$. We find, in units such that $\hbar = 1$, that DOS $\rho^{\sigma,\sigma}(k,\omega) = (1/2\pi)A^{\sigma,\sigma}(k,\omega)$. Upon using the result $(x \pm i\, 0^+)^{-1} = [P(x^{-1}) \pm (1/i)\pi\delta(x)]$, where P represents a Cauchy's principal value, we find that the DOS $\rho^{\sigma,\sigma}(k,\omega) = (1/2\pi)A^{\sigma,\sigma}(k,\omega)$ is given by a bunch of delta functions (a Fermi-liquid-like feature).

In this communication our aim is to establish first a relation between thermodynamic potential $\Omega = -(1/\beta)\ln \text{Tr} \exp(-\beta H)$ of the system and certain spectral functions. This is expected to yield an expression for entropy in closed form. The relation to be obtained is useful for deriving thermodynamics of the system in the pseudo-gapped state. The methodology followed is similar to that of Kadanoff and Baym[8]. About five decades ago these authors had established a formula relating thermodynamic average of a model Hamiltonian, for an interacting Bose system in the normal phase, to a spectral weight function. For the purpose stated, it is convenient to define a new thermodynamic potential $\Omega(\lambda)$ in terms of the Hamiltonian $H(\lambda) = \lambda H$ where $\lambda$ is a variable. For the pure DDW state without bi-layer splitting, to be discussed in this communication, H is given by

$$H'_{DDW} = \sum_{k\sigma} [\varepsilon_k d^\dagger_{k,\sigma} d_{k,\sigma} + \varepsilon_{k+Q} d^\dagger_{k+Q,\sigma} d_{k+Q,\sigma}]$$

$$+ \sum_{k\sigma} [(i\Delta_k) d^\dagger_{k,\sigma} d_{k+Q,\sigma} - (i\Delta_k) d^\dagger_{k+Q,\sigma} d_{k,\sigma}]. \qquad (6)$$

As shown above since the effect of bi-layer splitting is not of much significance, we do not include it in this exercise. The generalization, with the inclusion of the bi-layer splitting, is quite straightforward. For DDW-DSC state to be discussed in the sequel to this communication, however, the Hamiltonian H can be expressed as

$$H_{DDW-DSC} = H'_{DDW} + H_{DSC},$$

$$H_{DSC} = \sum_k [\Delta_{sc}(k) d^\dagger_{k,\uparrow} d^\dagger_{-k,\downarrow} + \Delta_{sc}(k) d_{-k,\downarrow} d_{k,\uparrow}] \qquad (7)$$

where $\Delta_{sc}(k) = \sum_{k'} U_1(k',k) \langle d_{-k',\downarrow} d_{k',\uparrow} \rangle$. We introduce additional anomalous pairing terms involving inter-site d-fermion interaction $U_1$ which correspond to the DSC component in the DDW-DSC order in the under-doped regime. In momentum space, the interaction $U_1$ assumed to be effective for nearest-neighbor(NN) only and corresponds to, say, $U_1(\mathbf{k'},\mathbf{k})$ for transition from a momentum $\mathbf{k'}$ to $\mathbf{k}$. The interaction $U_1(\mathbf{k'},\mathbf{k})$ is expressed in terms of basis function corresponding to $d_{x^2-y^2}$ state: we write $U_1(\mathbf{k'},\mathbf{k}) = g_1 \acute{\eta}_{1k'} \acute{\eta}_{1k}$ where $g_1$ is the coupling of the effective interaction in $d_{x^2-y^2}$ angular momentum state. In two dimensions we have $\acute{\eta}_{1k} = \acute{\eta}_{1k}(k_x, k_y)$, where for $d_{x^2-y^2}$ wave $\acute{\eta}_{1k} = (\cos k_x a - \cos k_y a)$. It was realized quite some time ago that a d-wave pairing has an advantage in that the electrons in a Cooper pair avoid each other, i.e. the pair wave function has zero amplitude at $\mathbf{r} - \mathbf{r'} = 0$, strongly reducing their local Coulomb repulsion. Thus, a contact Coulomb repulsion does not affect d-wave superconductivity. The d-wave symmetry implies that $U_1(k',k) = -U_1(k',k+Q)$ or $\Delta_{sc}(k+Q) = -\Delta_{sc}(k)$. The spin and charge pairings, namely

$$\Delta_s(k) = -\sum_{k'} U_1(k',k) \langle d^\dagger_{k',\sigma} d_{k'+Q,-\sigma} \rangle, \quad \Delta^\dagger_s(k) = -\sum_{k'} U_1(k',k) \langle d^\dagger_{k'+Q,-\sigma} d_{k',\sigma} \rangle$$

$$\Delta_c(k) = -\sum_{k'} U_1(k',k) \langle d^\dagger_{k',\sigma} d_{k'+Q,\sigma}\rangle, \quad \Delta^\dagger_c(k) = -\sum_{k'} U_1(k',k) \langle d^\dagger_{k'+Q,\sigma} d_{k',\sigma}\rangle \quad (8)$$

are though possible through the involvement of the interaction $U_1(k',k)$, will not be considered here. The absence of conventional spin or charge density wave orders in (7) play a crucial role in simplifying our analysis of the single particle excitation spectrum as will be seen in the forthcoming communication.

One can write $\Omega(\lambda) - \Omega(0) = \int (d\lambda/\lambda) \langle H(\lambda)\rangle_\lambda$ where $\Omega(0)$ is an integration constant and the angular brackets $\langle \ldots \rangle_\lambda$ denote thermodynamic average calculated with $H(\lambda)$. The system under consideration corresponds to $\Omega(\lambda=1)$. Obviously, the task now boils down to establishing relation between the average $\langle H(\lambda)\rangle_\lambda$ and spectral weight functions. The spectral functions for the pure DDW state are given by

$$A^\lambda(k_1,\sigma,\omega) = i\,[\,G^\lambda(k_1,\sigma,\omega_n)\big|_{i\omega_n = \omega + i0+} - G^\lambda(k_1,\sigma,\omega_n)\big|_{i\omega_n = \omega - i0+}\,], \quad (9)$$

$$G^\lambda(k_1,\sigma,k_1',\sigma',\omega_n) = \int_0^\beta d\tau\, e^{i\omega_n \tau} G^\lambda(k_1,\sigma,\tau; k_1',\sigma',0),\; k_1=(k,k+Q),\; k_1'=(k'+Q), \quad (10)$$

$$G^\lambda(k_1,\sigma,\tau; k_1',\sigma',\tau') = -\langle T\{d_{k_1,\sigma}(\tau)\,d^\dagger_{k_1',\sigma'}(\tau')\}\rangle_\lambda, \quad (11)$$

$$d_{k_1,\sigma}(\tau) = \exp(H(\lambda)\tau)\, d_{k_1,\sigma}\, \exp(-H(\lambda)\tau), \quad (12)$$

$i\omega_n = [(2n+1)\pi I/\beta]$ with $n = 0, \pm 1, \pm 2, \ldots \ldots$, and T is the time-ordering operator which arranges other operators from right to left in the ascending order of imaginary time $\tau$. The Lehmann representations (LR) of $G^\lambda(k_1,\sigma,\omega_n)$ can be obtained easily. We find

$$G^\lambda(k_1,\sigma,\omega_n) = \exp(\beta\Omega(\lambda)) \sum_{mn} e^{-\beta H_m(\lambda)} \langle m|\,d^\dagger_{k_1,\sigma}\,|n\rangle \langle n|\,d_{k_1,\sigma}\,|m\rangle$$
$$\times \{(1+ e^{\beta(H_m(\lambda)-H_n(\lambda))})/(i\omega_n + H_n(\lambda) - H_m(\lambda))\}. \quad (13)$$

Here $|m\rangle$ is an exact eigen state of $H(\lambda)$ and $\{H(\lambda)|m\rangle = H_m(\lambda)|m\rangle\}$. Upon using (13) in (9) we also obtain LR of $A^\lambda(k_1,\sigma,\omega)$:

$$A^\lambda(k_1,\sigma,\omega) = 2\pi\exp(\beta\Omega(\lambda))\sum_{mn} e^{-\beta H_m(\lambda)} \langle m| d^\dagger_{k,\sigma} |n\rangle \langle n| d_{k,\sigma} |m\rangle$$

$$\times (e^{\beta\omega} + 1)\,\delta(\omega + H_n(\lambda) - H_m(\lambda)). \qquad (14)$$

To fulfill the aim stated above, we now set up equations for the operators $\acute{O}(t) = d_{k,\sigma}(t)$, $d_{k+Q,\sigma}(t)$, etc. using the equation of motion (in units such that $\hbar = 1$) $i(\partial/\partial t)\acute{O}(t) = [\acute{O}(t), H(\lambda)]$, where $\acute{O}(t) = \exp(i H(\lambda)t)\,\acute{O}\,\exp(-iH(\lambda)t)$. With the help of these equations and Eq.(6) it is easy to see that

$$\langle H'_{DDW}(\lambda)\rangle_\lambda = (1/2)\,\mathrm{Lim}_{t'\to t}\sum_{k\sigma}\{(i/\lambda)(\partial/\partial t) - (i/\lambda)(\partial/\partial t')\}\{\langle d^\dagger_{k,\sigma}(t')d_{k,\sigma}(t)\rangle_\lambda$$

$$+ \langle d^\dagger_{k+Q,\sigma}(t')d_{k+Q,\sigma}(t)\rangle_\lambda\} \qquad (15)$$

for the pure DDW state. It is convenient to introduce the functions $f^\lambda(k,\sigma,\omega)$ and $f^\lambda(k+Q,\sigma,\omega)$, where $f^\lambda(k_1,\sigma,\omega) \equiv [-i\iint dt\,dt'\, e^{i\omega(t-t')} \langle d^\dagger_{k1,\sigma}(t')d_{k1,\sigma}(t)\rangle_\lambda \Theta(t-t')]$ and $\Theta(t)$ is the unit step-function given by $\Theta(t) = i\int_{-\infty}^{+\infty}(d\omega/2\pi)(e^{-i\omega t}/(\omega+i0^+))$, for our purpose. Here $0^+$ tends towards zero from positive values. The LR of $f^\lambda(k,\sigma,\omega)$ and $f^\lambda(k+Q,\sigma,\omega)$ can be obtained easily using this integral representation of $\Theta(t)$ and the identity $(x \pm i\,0^+)^{-1} = [P(x^{-1}) \pm (1/i)\pi\delta(x)]$ valid for real $\omega$, where $P$ represents a Cauchy's principal value. We find

$$\mathrm{Im}\, f^\lambda(k_1,\sigma,\omega) = -\pi\exp(\beta\Omega(\lambda))\sum_{mn} e^{-\beta H_m(\lambda)} \langle m| d^\dagger_{k1,\sigma} |n\rangle \langle n| d_{k1,\sigma} |m\rangle$$

$$\times \delta(\omega + H_n(\lambda) - H_m(\lambda)), \qquad (16)$$

and

$$\mathrm{Re}\, f^\lambda(k_1,\sigma,\omega) = -P\int_{-\infty}^{+\infty}(d\omega'/\pi)\,\{\mathrm{Im}\, f^\lambda(k_1,\sigma,\omega)/(\omega-\omega')\}. \qquad (17)$$

Upon comparing (12) with (10) we obtain $\mathrm{Im}\, f^\lambda(k_1,\sigma,\omega) = (-1/2)(e^{\beta\omega} + 1)^{-1} A^\lambda(k_1,\sigma,\omega)$. In view of this result and Eq.(13) one can write

$$f^\lambda (k_1,\sigma,\omega) = [-i \iint dt\, dt' \int_{-\infty}^{+\infty}(d\omega'/2\pi)e^{i(\omega-\omega')(t-t')}( e^{\beta\omega'} + 1)^{-1} A^\lambda (k_1,\sigma,\omega') \Theta(t-t')] \quad (18)$$

which immediately yields $\langle d^\dagger_{k_1,\sigma}(t')d_{k_1,\sigma}(t)\rangle_\lambda = \int_{-\infty}^{+\infty}(d\omega'/2\pi)e^{-i\omega'(t-t')}(e^{\beta\omega'} + 1)^{-1} A^\lambda (k_1,\sigma,\omega')$ and the relation sought for, viz.

$$\Omega(\lambda) - \Omega(0) = \int (d\lambda/\lambda)\sum_{k\sigma} \int_{-\infty}^{+\infty}(d\omega/2\pi)\,\omega\,(e^{\beta\omega} + 1)^{-1}\{A^\lambda(k,\sigma,\omega) + A^\lambda(k+Q,\sigma,\omega)\}. \quad (19)$$

The straightforward method of calculating these weight functions comprises of setting up equations of motion for the temperature functions $G^\lambda(k_1,\sigma,\tau; k_1',\sigma',\tau')$ and then obtain the corresponding Matsubara propagators $G^\lambda(k_1,\sigma, k_1',\sigma',\omega_n)$. The weight functions ($A^\lambda(k,\sigma,\omega)$, $A^\lambda(k+Q,\sigma,\omega)$) can then be obtained using Eq.(5).

For the pure DDW state we find that the weight functions $A^\lambda (k,\sigma,\omega) = 2\pi[ u_k^2 \delta(\omega - \lambda E_k^U) + v_k^2 \delta(\omega - \lambda E_k^L)]$ and $A^\lambda (k+Q,\sigma,\omega) = 2\pi[ v_k^2 \delta(\omega - \lambda E_k^U) + u_k^2 \delta(\omega - \lambda E_k^L)]$ where $u_k^2 = (1/2)[1+ (\varepsilon_k^L/((\varepsilon_k^L)^2 + \Delta_k^2)^{1/2})]$ and $v_k^2 = (1/2)[1- (\varepsilon_k^L/((\varepsilon_k^L)^2 + \Delta_k^2)^{1/2})]$. The single particle excitation spectrum $E_k^{(U,L)} = \varepsilon_k^U \pm [(\varepsilon_k^L)^2 + \Delta_k^2]^{1/2}$ where $\varepsilon_k^U = (\varepsilon_k + \varepsilon_{k+Q})/2$ and $\varepsilon_k^L = (\varepsilon_k - \varepsilon_{k+Q})/2$. This eventually yields the expression for the thermodynamic potential per unit cell as a function of $\lambda$: $\Omega_{DDW}(\lambda) = \int d\lambda\, (N_s)^{-1}\sum_k [E_k^U(1-\tanh(\beta\lambda E_k^U/2)) + E_k^L(1-\tanh(\beta\lambda E_k^L/2))]$ where $N_s$ is the number of unit cells and $\beta = (k_B T)^{-1}$. For the real system, in the absence of magnetic field, one can write

$$\Omega_{DDW}(\lambda=1) = \Omega_0 - 2(\beta N_s)^{-1}\sum_{k,j=U,L}\{\ln\cosh(\beta E_k^{(j)}/2)\} \quad (20)$$

where $\Omega_0 = (N_s)^{-1}\sum_{k,j=U,L} E_k^{(j)}$. The dimensionless entropy per unit cell is given by $s = -(\partial\Omega/\partial(k_B T)) = \beta^2(\partial\Omega/\partial\beta)$. We obtain for the pseudo-gapped phase

$$s_{DDW} = (2/N_s)\sum_{k,j=U,L}[\ln(1/2) + \ln(1+\exp(-\beta E_k^{(j)}))$$
$$+(\beta E_k^{(j)} + \beta^2(\partial E_k^{(j)}/\partial\beta))(\exp(\beta E_k^{(j)})+1)^{-1}]. \quad (21)$$

It is now easy to see that in the pseudo-gapped phase, with bi-layer splitting, the entropy will be given by the generically same expression as in (21); the summation over j will, however, include j = $(U^{\pm}, L^{\pm})$. In the normal (non-pseudo-gapped) phase, the dimensionless entropy is given by

$$s_{normal} = (2/N_s) \sum_{k,j=1,2} [\ln(1/2) + \ln(1+\exp(-\beta\,\varepsilon^{(j)}(k)))$$
$$+ (\beta\varepsilon^{(j)}(k) + \beta^2(\partial\varepsilon^{(j)}(k)/\partial\beta))(\exp(\beta\varepsilon^{(j)}(k))+1)^{-1}] \qquad (22)$$

where $\varepsilon^{(1)}(k) = \varepsilon_k$, and $\varepsilon^{(2)}(k) = \varepsilon_{k+Q}$. The specific heat per unit cell can now be formally expressed as $c = -\beta(\partial s/\partial\beta)$. At 10% doping level, in the vicinity of T*, upon treating the gap $\Delta_k^2$ as a small parameter and ignoring the dependence of chemical potential on $\beta$ we are able to show that ($F(\beta^*) \equiv \beta^{*2} \exp(\beta\varepsilon^{(j)}(k))(\exp(\beta\varepsilon^{(j)}(k))+1)^{-2}$)

$$[s_{DDW}(T<T^*) - s_{normal}(T>T^*)] = (2/N_s)\sum_{k,j=1,2} \{F(\beta^*)\Delta_k^4(T<T^*)/8(\varepsilon_k^L)^2\}$$
$$= (\beta^{*2}\Delta_0^4(T<T^*)/1024t^2 N_s) \sum_{k,j=1,2} (\acute{\eta}_k^4/\zeta_k^2) \operatorname{sech}^2(\beta^*\varepsilon^{(j)}(k)/2) \qquad (23)$$

where $\acute{\eta}_k = (\cos k_x a - \cos k_y a)$, and $\zeta_k = (\cos k_x a + \cos k_y a)$. The upper/lower limits of the momentum integral in (23) has been determined from the FS reconstruction in Fig.1. We find the right-hand-side equal to $0.5494(\beta^{*2}\Delta_0^4(T<T^*)/1024t^2) > 0$. Thus the pseudo-gap(PG) transition is a first order one and the entropy per unit cell increases in the hidden-order state.

In the presence of magnetic field **B**, the situation is different. Suppose the Cu-O plane of the system is the x-y plane. For a magnetic field applied in z-direction (i.e. the vector potential $\mathbf{A} = (0, -Bx, 0)$ in Landau gauge), starting with a tight binding model, it can be easily seen that the band energies $E_k^{(j,\nu)}(k) = \varepsilon_k^U + \nu\,t_\perp(k) + j[(\varepsilon_k^L)^2 + \Delta_k^2]^{1/2}$ are replaced by $E_k^{(j,\nu)}(B)$ given by $E_k^{(j,\nu)}(B) = \varepsilon_k^U(B) + \nu\,t_\perp(k) + j[(\varepsilon_k^L(B))^2 + \Delta_k^2]^{1/2}$. Upon considering the hopping processes upto the second neighbor only, one may write $\varepsilon_k^U(B) = (\varepsilon_k(B) + \varepsilon_{k+Q}(B))/2$ and $\varepsilon_k^L(B) = (\varepsilon_k(B) - \varepsilon_{k+Q}(B))/2$ where

$$\varepsilon_k(B) = -\mu' - 2t(\cos(k_x a) + \cos(k_y a + \varphi)) + 4t'\cos(k_x a)\cos(k_y a + \varphi/2), \quad (24)$$

and

$$\mu' \equiv \mu - \hbar \sum_{n=0}^{\infty} (2n+1)(\omega_c/2) + (-1)^{\sigma}(g\mu_B B/2), \quad n = 0, 1.... \quad (25)$$

The first term in (25) includes the chemical potential $\mu$, the Landau levels and cyclotron frequency $\omega_c = eB/m^*$ ($m^*$ is the effective mass of the electrons), and the Zeeman term ($g\mu_B B/2$). The quantity $\varphi = (2\pi eBa^2/h)$ is the Peierls phase factor[6,7] and 'a' is the lattice constant. Equation (23) enables one to analyze the density of state (DOS) and electron density $n_e(B,T)$ in the presence of magnetic field. As in Eq.(2), ignoring the Zeeman term, we find that $n_e(B,T) = 1 - (N_s)^{-1}\sum_{k\sigma} n(k,B,T)$ where $n(k,B,T) = \sum_{j=\pm}\{u_k^{(j)2}(B)(\exp(\beta E_k^{U,j}(B))+1)^{-1} + v_k^{(j)2}(B)(\exp(\beta E_k^{L,j}(B))+1)^{-1}\}$. This gives

$$n_e(B,T) = (N_s)^{-1}\sum_{k,j=\pm}\{u_k^{(j)2}(B)\tanh(\beta E_k^{U,j}(B)/2) + v_k^{(j)2}(B)\tanh(\beta E_k^{L,j}(B)/2)\}. \quad (26)$$

Now with increase in B, the Landau states move to a higher energy, ultimately rising above the Fermi level. They are thereby emptied, and the excess electrons find a place in the next lower Landau level(LL). During the crossing of a LL, its occupation by electrons is halted and then reduced, the electrical conductivity consequenty decreases slightly. As the excess electrons get accommodated in the next lower LL, the conductivity rises again. These oscillations in the electron density in the vicinity of Fermi energy manifest themselves as oscillations in electrical conductivity (Shubnikov-de Haas oscillations (SdHO)). In YBCO these oscillations are believed to have their origin mainly at electron pockets in Fermi surface located around the anti-nodal points. Thus the magnetic field dependent occupancy $n_e([(\pm\pi, 0), (0,\pm\pi)],B,T)$ at these points are expected give a reasonable idea about the frequencies of these oscillations. With this hope, in Fig. 2, we have plotted the occupancy at the anti-nodal points as a function of magnetic field for T = 1 K at 10% doping level. From this figure we find that the halting, alluded to above, occurs at $B_1 \sim 350$ Tesla. We also find that, at T ~30 K, the halting gets eroded (see Fig.3).The previous workers [6,7] have reported that there are two main oscillation frequencies $B_1 = 500 \pm 30T$ and $B_2 = 910 \pm 30T$, corresponding to the electron and hole pockets respectively; the value of $B_1$ agrees with experimental observations ( the second frequency $B_2$ has not been observed yet). It is, therefore, satisfactory to observe that the crude estimate of $B_1$ presented here is not off the mark and SdHO (in YBCO) does not occur at higher temperatures (T > 30 K). In a future communication we plan to calculate the electrical conductivity tensor components to estimate these values accurately.

Thermoelectricity as a probe of the pseudo-gapped state is still largely unexplored. In what follows we discuss this issue briefly. When an electric field **E** is applied within the conducting plane of the system and a magnetic field **B** in the perpendicular direction, a quasi-particle in DDW drifts with velocity $\mathbf{v}_D$ perpendicular to both **B** and **E** ($\mathbf{v}_D = (\mathbf{E} \times \mathbf{B})/B^2$). The heat current parallel to $\mathbf{v}_D$ is then given by $\mathbf{J}_H = \beta^{-1} S_{DDW} \mathbf{v}_D$, where $S_{DDW}$ is the total entropy associated with the quasi-particles in DDW. In fact, at **B** = 0, the electric field tends to become orthogonal to the thermal current. The presence of magnetic field takes away this alignment. The quantity $S_{DDW} = s_{Landau} + s_{DDW}(B)$ where

$$s_{Landau} = \sum_n [\ \ln(1+\exp(-\beta E_n)) + \beta E_n (\exp(\beta E_n)+1)^{-1}\ ], \tag{27}$$

$E_n = (n+(1/2))\ \hbar \omega_c$ (where n = 0,1,2,…), and $\omega_c = eB/m^*$. The sum in (27) has to be taken over all the Landau levels. For small T and large B, Eq. (27) is well approximated by taking the n = 0 and n = 1 Landau levels. The entropy $s_{DDW}(B)$ is obtained as in Eq.(21).

The Nernst effect corresponds to the off diagonal component of the thermoelectric power in the presence of magnetic field. In the configuration discussed above, the Nernst coefficient $S_{xy} = - (k_B\, S_{DDW}\, \rho/B)$ μV/K calculation appears to be possible. Here $\rho$ is the magneto-resistivity per unit cell volume within the two level approximation. The opening of a d-wave pseudo-gap(in high- $T_c$ cuprates) lowers the carrier density as the gap destroys much of the Fermi surface. This restricts the phase space and leads to an increase in the mean free-path of the nodal quasi-particles. These features are expected to conspire in such a way so as to create a large Nernst effect. A theoretical investigation of this effect is one of the objectives of a sequel to this communication.

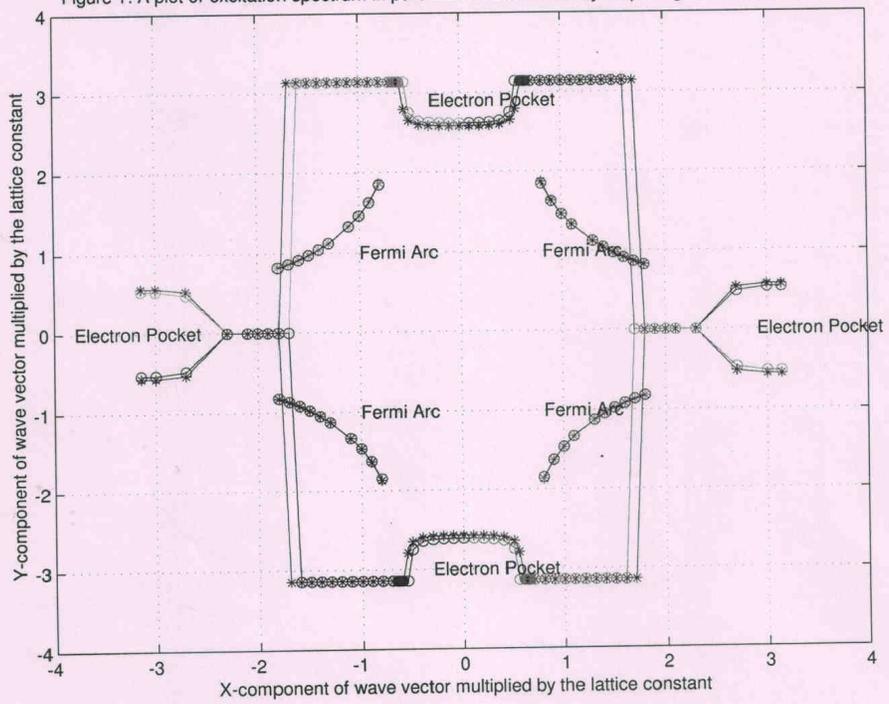

Figure 1: A plot of excitation spectrum in pure DDW state with bilayer splitting for 10% hole doping.

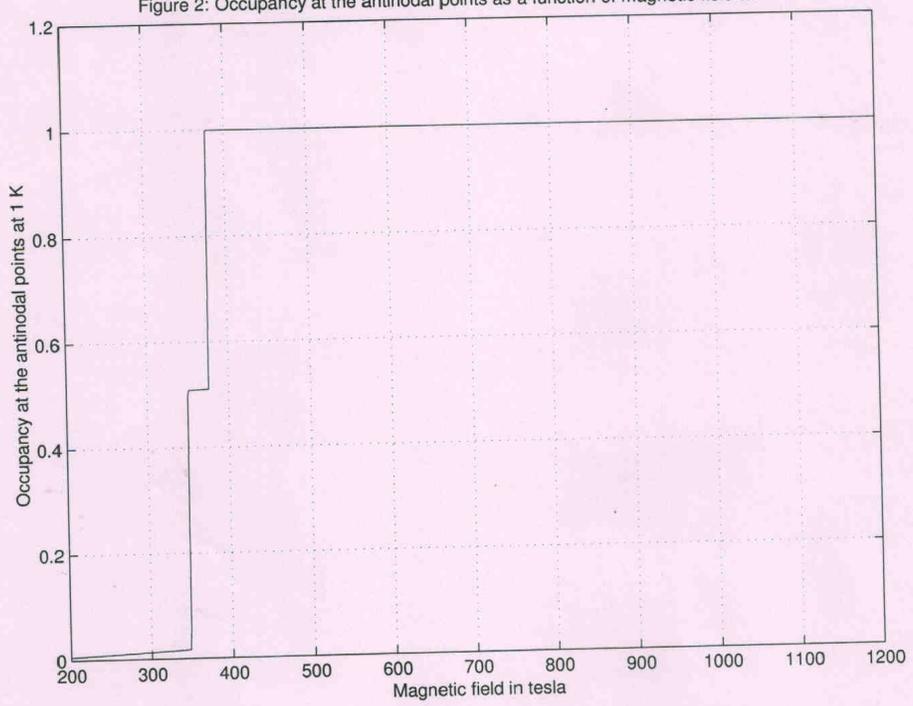

Figure 2: Occupancy at the antinodal points as a function of magnetic field at 1 K.

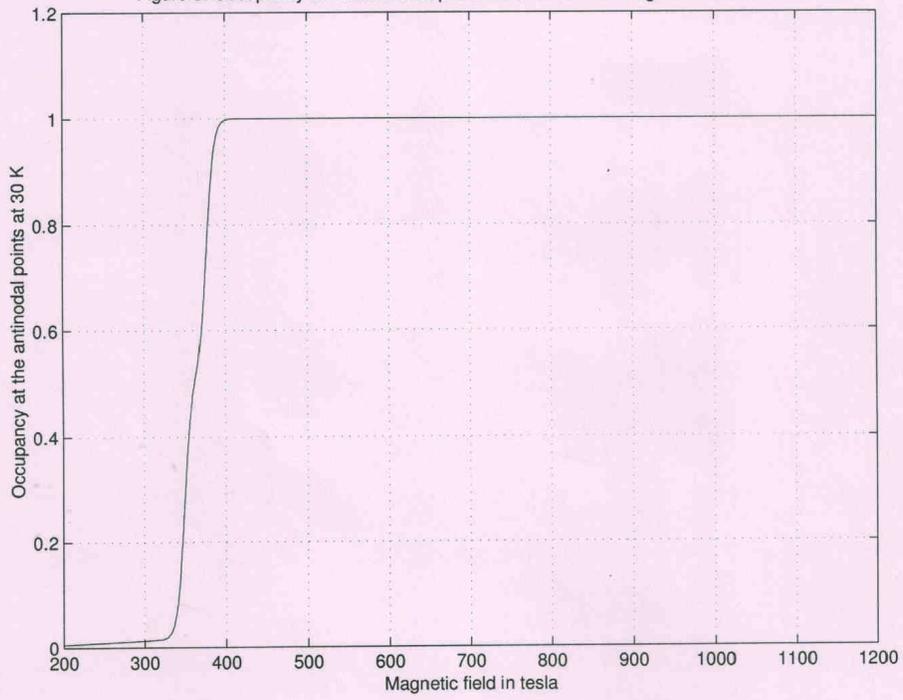

Figure 3: Occupancy at the antinodal points as a function of magnetic field at 30 K.